\documentclass[a4paper,11pt]{article}
\usepackage{jheppub} 
\usepackage{lineno}
\usepackage{amsfonts}
\usepackage{latexsym}
\usepackage{amsmath,amssymb}
\usepackage{verbatim}
\usepackage{setspace}
\usepackage{color}
\usepackage{physics}
\usepackage{tikz}
\usepackage{mathdots}
\usepackage{yhmath}
\usepackage{cancel}
\usepackage{color}
\usepackage{siunitx}
\usepackage{mathrsfs}  
\usepackage{array}
\usepackage{multirow}
\usepackage{amssymb}
\usepackage{gensymb}
\usepackage{tabularx}
\usepackage{hyperref}
\usepackage{booktabs}
\numberwithin{figure}{section}
\usetikzlibrary{fadings}
\usetikzlibrary{patterns}
\usetikzlibrary{shadows.blur}
\usetikzlibrary{shapes}

\numberwithin{equation}{section}

\newcommand{\bea}{\begin{eqnarray}}
\newcommand{\eea}{\end{eqnarray}}
\newcommand{\be}{\begin{equation}}
\newcommand{\ee}{\end{equation}}
\newcommand{\ba}{\begin{align}}
\newcommand{\ea}{\end{align}}

\renewcommand{\epsilon}{\varepsilon}

   \makeatletter
  \let\over=\@@over \let\overwithdelims=\@@overwithdelims
  \let\atop=\@@atop \let\atopwithdelims=\@@atopwithdelims
  \let\above=\@@above \let\abovewithdelims=\@@abovewithdelims
\renewcommand\section{\@startsection {section}{1}{\z@}%
                                   {-3.5ex \@plus -1ex \@minus -.2ex}
                                   {2.3ex \@plus.2ex}%
                                   {\normalfont\large\bfseries}}

\renewcommand\subsection{\@startsection{subsection}{2}{\z@}%
                                     {-3.25ex\@plus -1ex \@minus -.2ex}%
                                     {1.5ex \@plus .2ex}%
                                     {\normalfont\bfseries}}

\arxivnumber{2407.12521} 

\title{\boldmath Marginality from Leading Soft Gluons}

\author[a]{Sruthi A. Narayanan}
\affiliation[a]{Perimeter Institute for Theoretical Physics,\\
Waterloo, ON N2L 2Y5, Canada
}

\emailAdd{sruthi81294@gmail.com}

\abstract{Marginal operators in a $d$-dimensional conformal field theory (CFT), those with conformal dimension $\Delta=d$, give us information about the space of related theories. This can be incredibly useful when trying to develop an intrinsic definition of a holographic CFT. Recent work has established that in the case of celestial CFTs, those dual to asymptotically flat spacetimes, marginal operators can be constructed as shadow transforms of soft scalars. In this work, rather than starting from a bulk theory of scalars, we start with Yang-Mills and construct marginal operators from soft gluons via light transforms. We obtain the local metric and connections for the associated conformal manifold by studying the two- and three-point functions of these marginal operators. We then discuss the implications of this for broadening our understanding of operator product expansions of soft operators in celestial CFTs.}

\begin{document}
\maketitle
\flushbottom

\section{Introduction}
The defining ingredients of a conformal field theory (CFT) are the spectrum of operators, i.e the operators $\mathcal{O}_i$ and their conformal dimensions $\Delta_i$, as well as the operator product expansion (OPE) coefficients. Given this information, one can completely specify the theory since higher point correlators can be built up from lower point correlators by way of the OPE.\cite{DiFrancesco:639405} However, generally this information is difficult to collect for a given CFT, especially when there is not an intrinsic Lagrangian construction. Celestial CFTs, those that are dual to asymptotically flat spacetimes, possess the advantage that their fundamental information is derived from bulk scattering amplitudes.\cite{Raclariu:2021zjz, Pasterski:2021rjz} Since bulk Lorentz symmetry is manifest as two-dimensional (2D) conformal symmetry on the boundary two-sphere (the celestial sphere) one can transform $S$-matrix elements to 2D objects that have the behavior of conformal correlators. These are often referred to as celestial amplitudes and are the fundamental building blocks of celestial holography. Likewise, the collinear behavior of scattering amplitudes gives rise to the OPE data in celestial CFTs by way of collinear splitting functions.\cite{Pate:2019lpp}

However, deriving properties of celestial CFTs from bulk scattering amplitudes is naturally constrained by the knowledge we have about the amplitudes themselves. Therefore, it seems that we will only know as much about celestial CFTs as we know about our bulk theory. Since one of the goals of the celestial holography program is to use this duality to learn about a putative bulk theory of quantum gravity, one is motivated to develop a more intrinsic understanding of what a celestial CFT is. One way to do this is to study deformations to the boundary theory. In~\cite{Ball:2021tmb,Melton:2022fsf,Melton:2023lnz}, the authors studied deformations arising from adding loop corrections and also at tree level by adding in a background field in the bulk theory. However, this still involved transforming the modified bulk amplitudes and observing that effect on celestial CFT correlators. 

An approach that is relatively agnostic to the bulk, is the one taken in~\cite{Kapec:2022axw, Kapec:2022hih} where the authors studied the addition of marginal operators $\mathcal{M}_I$ in the CFT. Marginal operators in two dimensions are operators with conformal dimension $\Delta = 2$ and 2D spin $J=0$, i.e they transform like scalars. As opposed to relevant $(\Delta<2)$ and irrelevant $(\Delta>2)$ operators, the addition of a marginal operator into the spectrum preserves the conformal symmetry of the theory and therefore gives rise to a distinct but equally well-defined CFT. The standard approach to understanding deformations by a marginal operator is to start with an action $S$ and then deform it by a term of the form $\delta S = \lambda^I\int d^2 x \mathcal{M}_I$. For each marginal operator $\mathcal{M}_I$, there is an associated parameter $\lambda^I$ and each distinct set of $\lambda^I$'s will give rise to a new CFT. We can think of this as a manifold where each point corresponds to a distinct CFT. The marginal operators serve as directions one can travel on this manifold and the $\lambda^I$'s determine how much one travels in each direction.\cite{Balthazar:2022hzb} 

The beauty of this conformal manifold, is that it comes equipped with an intrinsic geometry that can be used to understand the full space of possibly related CFTs. In~\cite{Kapec:2022axw} the authors explored the connections between a bulk non-linear sigma model (NLSM) and the geometry of the manifold of associated dual celestial CFTs. Due to the relationship between the manifold and the geometric soft theorems given in~\cite{Cheung:2021yog} there was a nice interpretation in terms of the target manifold of the NLSM. The marginal operators generating movement on the conformal manifold were given by shadow transformations of soft scalars. In~\cite{Kapec:2022hih} a somewhat analogous connection was made in the case of bulk gauge and gravity theories. However, the standard shadow transform shifts the left and right conformal weights by $(h,\bar{h})\rightarrow(1-h,1-\bar{h})$ which is equivalent to saying that the dimension and 2D spin are shifted according to $(\Delta,J)\rightarrow (2-\Delta,-J)$. Therefore, while shifting a dimension $\Delta=0$ soft operator will give an operator of dimension $\Delta=2$, that operator will only be marginal if it is a scalar since a marginal operator must have 2D spin $J=0$. As such, it seems like the operators considered in~\cite{Kapec:2022hih} are not exactly marginal operators but rather some kind of ``quasi-marginal" operators that can be nicely related to the soft theorems.

The OPE of these ``quasi-marginal" operators will not, in the traditional~\footnote{For the shadowed soft operators in gauge theory and gravity, there probably is a manifold where each point consists of a CFT with a different copy of the conformal group. There should be a method to obtain the geometric aspects, however it is most not from reading off terms in the OPE.} sense, encode aspects of the geometry of the conformal manifold. To that end, to find the geometry from the OPE one needs to construct operators that have the right conformal weights. If the spectrum of operators contains those of integer conformal dimension~\cite{Freidel:2022skz, Cotler:2023qwh}, when the operators have 2D spin $J=\pm 1,\pm 2$, none of these can be marginal and we must consider some appropriate transformations of these operators. Since it does not seem plausible to do so equipped with just the shadow transformation, in this paper we will explore the transforms that generate operators that have the right conformal weights to be marginal. In particular, we will use the light transform. The light transform has appeared in much of the recent literature on celestial holography~\cite{Guevara:2021abz,Strominger:2021mtt,Himwich:2021dau} as it has allowed for a clean understanding of the boundary soft algebras. 

This paper is organized as follows. In section~\ref{sec:conventions} we will review the necessary conventions for the remainder of the paper including the definition of celestial amplitudes, soft operators and some aspects of marginal operators. In section~\ref{sec:casimir} we discuss the transformations that preserve the conformal casimir and form the group $D_8$. We find that in two dimensions, there are only four relevant transformations: the identity, shadow, left and right light transforms. In section~\ref{sec:marginalops} we use the light transforms to construct marginal operators from the leading soft gluons in Yang-Mills. In sections~\ref{sec:metric} and~\ref{sec:connections} we use the two- and three-point functions of these marginal operators to construct the metric and connections on the associated conformal manifold. We are then able to deduce the form of the marginal OPEs and in section~\ref{sec:discussion} we speculate on how we could use this data to learn more about the intrinsic structure of celestial CFTs.

\section{Conventions}\label{sec:conventions}
In this section we will outline the definitions and background necessary for our results. We will work in flat Minkowski space which has the following metric in four dimensions and $(3,1)$ signature
\begin{equation}
    ds^2 = -dx_0^2+dx_1^2+dx_2^2+dx_3^2.
\end{equation}
A null vector $q^\mu$ can be parametrized by 
\begin{equation}
    q^\mu(\omega,z,\bar{z}) = \eta\omega \hat{q}^\mu(z,\bar{z})  = \eta\omega (1+z\bar{z},z+\bar{z},-i(z-\bar{z}),1-z\bar{z})
\end{equation}
where $\omega$ is the energy, $\eta = +1\ (-1)$ for outgoing (incoming) particles, and $(z,\bar{z})$ labels a coordinate on the celestial sphere. Since we are working in (3,1) signature, $z$ and $\bar{z}$ are complex conjugates of one another however, we can generally analytically continue to $(2,2)$ signature Klein space~\cite{Atanasov:2021oyu} where $z,\bar{z}$ are real and independent. 

Scattering amplitudes in 4D asymptotically flat spacetimes can be recast as correlation functions on the 2D boundary that transform appropriately under conformal transformations. These boundary conformal correlators are often referred to as celestial amplitudes. Celestial amplitudes are found by computing scattering amplitudes in a basis that diagonalizes boosts rather than translations. For massless external particles, one transforms the scattering amplitude by Mellin transforming the momentum space scattering amplitude with respect to the energies of each particle
\begin{equation}
\mathcal{A}\left(1^{\eta_1,a_1}_{\Delta_1}\cdots n^{\eta_n,a_n}_{\Delta_n}\right) = \prod_{j=1}^n\left[\int_0^\infty d\omega_j\omega_j^{\Delta_j-1}\right]A_{a_1\ldots a_n}(\eta_1\omega_1\hat{q}(z_1,\bar{z}_1),\ldots,\eta_n\omega_n\hat{q}(z_n,\bar{z}_n))
\end{equation}
where $\Delta_n$ are the conformal weights of the external particles and $a_n$ is any other label, like helicity or color. We will write these amplitudes as correlation functions 
\begin{equation}
\mathcal{A}\left(1^{\eta_1,a_1}_{\Delta_1},\ldots n^{\eta_n,a_n}_{\Delta_n,a_n}\right) = \left\langle \mathcal{O}^{\eta_1,a_1}_{\Delta_1}\cdots \mathcal{O}^{\eta_n,a_n}_{\Delta_n}\right\rangle .
\end{equation} 

There has been significant recent work to try to understand the ``correct" basis of operators to work with in celestial CFTs. It is widely accepted that the set of operators with integer dimensions is a good basis to work in. This includes, as a subset, the soft operators, the insertions of which generate the standard soft theorems. As such, one can define the soft gluon operators~\cite{Guevara:2021abz}
\begin{equation}\label{eq:softgluons}
    R^{k,a,\pm}(z,\bar{z}) = \lim_{\varepsilon\to 0}\varepsilon\mathcal{O}^{\pm, a}_{k+\varepsilon}(z,\bar{z}),\ k = 1, 0, -1,\ldots 
\end{equation}
The correlators of these soft operators will be finite, as the vanishing $\varepsilon$ cancels a divergence arising from the small $\omega$ region of the Mellin transform. Alternatively, one can understand the right hand side of this equation as the residue~\cite{Freidel:2022skz} of the operator $\mathcal{O}_{\Delta}^{\pm, a}$ when $\Delta$ is a negative integer. We will be primarily concerned with the \textbf{leading soft gluon} operator which is the $k=1$ operator above. In what follows, whenever we refer to a soft gluon operator we implicitly mean the leading soft operator above and will omit the $k$ index to write it as $R^{\pm,a} \equiv R^{k=1,a,\pm}$. 

Lastly, it will be useful to review some important aspects of marginal operators in relation to the geometry of the conformal manifold. In two dimensions the OPE of two marginal operators has the following form~\cite{Kutasov:1988xb}
\begin{equation}\label{eq:margOPE}
\mathcal{M}_I(x)\mathcal{M}_J(y)\sim \frac{g_{IJ}}{|x-y|^4} + \Gamma_{IJ}^K \mathcal{M}_K(y)\delta^{(2)}(x-y)+\cdots
\end{equation}
The more regular terms will be proportional to the curvature tensor and its derivatives. Additionally, one can determine the connections by integrating the three point function of marginal operators 
\begin{equation}\label{eq:intthreept}
\frac{\partial g_{IJ}}{\partial\lambda^K} = \int d^2 z\langle \mathcal{M}_K(z)\mathcal{M}_I(1)\mathcal{M}_J(0)\rangle
\end{equation}
and then taking the usual combination of these to obtain 
\begin{equation}\label{eq:gammas}
\Gamma^K_{IJ} = \frac{1}{2}g^{KL}\left(\frac{\partial g_{LI}}{\partial\lambda^J}+\frac{\partial g_{LJ}}{\partial\lambda^I} - \frac{\partial g_{IJ}}{\partial\lambda^L}\right).
\end{equation}
It should be noted that the metric and connections are functions of the couplings $\lambda^i$. However, when we extract this information, it is the local data and so it should be understood as $g_{IJ}(0)$ and $\Gamma_{IJ}^K(0)$. In what follows we will not explicitly write that the argument is $0$. Additionally, when considering exactly marginal operators, it is always possible to choose local coordinates~\cite{Kutasov:1988xb} such that the connections vanish. In this work we do not pick these coordinates because the connections, while coming from distributional terms in the OPE, tell us information about the OPE of non-marginal operators in the celestial CFT which we do care about. We will discuss this in more detail later on.

Additionally, the integrand in~\eqref{eq:intthreept} should vanish when $z\neq 0,1$ if the operators in question are truly marginal. This is basically equivalent to saying that the three point function should evaluate to a contact term. In what follows we will determine the set of possible marginal operators that can be constructed from the Yang-Mills soft currents. All of the non-zero two-point functions will tell us the metric components from which we can try to understand the structure of the conformal manifold.

\section{Casimir Preserving Integral Transforms}\label{sec:casimir}
Since marginal operators are scalars (2D spin-$0$), as we mentioned previously, unless we start with a soft scalar the soft operators discussed above cannot directly be identified as marginal operators in the CFT. In the general case, we need to transform the soft operators into single particle operators that can be identified as marginal. Thus far, the only integral transforms encountered in the celestial CFT literature are the shadow and the light transforms.\cite{Pasterski:2017kqt,Sharma:2021gcz,De:2022gjn,Hu:2022syq,Banerjee:2024hvb} In this section we will discuss the complete set of transformations that preserve the quadratic Casimir and conclude that there are no additional relevant transformations.

In two dimensions, the allowed transforms are those that preserve the quadratic Casimir\footnote{In higher dimensions, one would need to preserve the higher Casimirs as well.}
\begin{equation}
C_2(\Delta,J) = \Delta(\Delta-2)+J^2 = 2\left(h\left(h-1\right)+\bar{h}\left(\bar{h}-1\right)\right).
\end{equation}
Just by looking at this expression, we can deduce transformations in the weights that will leave it invariant
\begin{equation}
(h,\bar{h})\rightarrow \left\{\underset{\mathcal{T}_1}{(h,\bar{h})},  \underset{\mathcal{T}_2}{(\bar{h},h)},  \underset{\mathcal{T}_3}{(1-h,\bar{h})},  \underset{\mathcal{T}_4}{(h,1-\bar{h})},  \underset{\mathcal{T}_5}{(1-\bar{h},h)}, \underset{\mathcal{T}_6}{(\bar{h},1-h)}, \underset{\mathcal{T}_7}{(1-h,1-\bar{h})}, \underset{\mathcal{T}_8}{(1-\bar{h},1-h)}\right\}.
\end{equation}
We have labeled each of the eight transformations by $\mathcal{T}_i$. In what follows, we will explicitly discuss the forms of these transformations.

We notice that $\mathcal{T}_1$ is obviously the identity, $\mathcal{T}_3, \mathcal{T}_4$ are the light transforms and $\mathcal{T}_7$ is the shadow, all of which are given by\footnote{Note that the normalizations for the light transforms are different from previous literature~\cite{Himwich:2021dau,Sharma:2021gcz}. Previously they were given with unit normalization however that was not consistent with the statement that the product of left and right light transforms is the shadow since that would imply a unit normalization for the shadow as well.} 
\begin{eqnarray}
\mathcal{T}_1 & = & \mathbb{I}\left[\mathcal{O}\right](w,\bar{w}) = \int d^2z \delta^{(2)}(z-w)\mathcal{O}(z,\bar{z})\cr
\mathcal{T}_3 &\equiv& \mathcal{L}^+\left[\mathcal{O}\right](w,\bar{w})  = \frac{\Gamma(2-2h)}{\sqrt{\pi}} \int dz\frac{1}{(w-z)^{2-2h}}\mathcal{O}_{h,\bar{h}}(z,\bar{w})\cr
\mathcal{T}_4 &\equiv& \mathcal{L}^-\left[\mathcal{O}\right](w,\bar{w})  = \frac{1}{\sqrt{\pi}\Gamma(2\bar{h}-1)}\int d\bar{z}\frac{1}{(\bar{w}-\bar{z})^{2-2\bar{h}}}\mathcal{O}_{h,\bar{h}}(w,\bar{z})\cr
\mathcal{T}_7 &\equiv& \mathcal{S}[\mathcal{O}](w,\bar{w})  =  \frac{\Gamma(2-2h)}{\pi\Gamma(2\bar{h}-1)}\int d^2z\frac{1}{(w-z)^{2-2h}(\bar{w}-\bar{z})^{2-2\bar{h}}}\mathcal{O}_{h,\bar{h}}(z,\bar{z}).
\end{eqnarray}
We note that the shadow is also the composition of a left and right light transform i.e $\mathcal{S} = \mathcal{L}^+\mathcal{L}^-$.

In principal, there should be four other transformations. These transformations were, in fact, enumerated in~\cite{2018JHEP...11..102K} for CFTs in arbitrary dimensions. However, in 2D the remaining four transforms can be obtained from these four by a discrete transformation that interchanges the conformal weights. To understand this explicitly, let us define a ``parity" operator $\mathcal{P}$ such that
\begin{equation}
\mathcal{P}\left[\mathcal{O}_{h,\bar{h}}\right] \equiv \mathcal{O}_{\bar{h},h}.
\end{equation}
Since the 2D coordinates $z,\bar{z}$ are complex conjugates of one another, one can also think of this as a holomorphic to antiholomorphic switch, i.e $z\leftrightarrow \bar{z}$ or a switch between left-movers and right-movers. It is important to note that since this is a \textbf{discrete} transformation, its action on an operator does not result in an independent operator. It is not difficult to see that the remaining four transformations are compositions of the above with $\mathcal{P}$. We enumerate this\footnote{We especially thank Sabrina Pasterski for suggesting the introduction of the parity operator $\mathcal{P}$ as it greatly helped with the understanding of these transformations in two dimensions.} in Table~\ref{tab}. 
\begin{table}[htb]
\begin{center}
\begin{tabular}{ |c|c|c|c|c| } 
 \hline
 Transform & $(\Delta',J')$  & $(h',\bar{h}')$ & Order &Comparison to~\cite{2018JHEP...11..102K} \\ 
 \hline
 $\mathbb{I}$ & $(\Delta,J)$ & $(h,\bar{h})$  &1& 1 (Identity)\\
 $\mathcal{L}^+$ & $(1-J,1-\Delta)$ & $(1-h,\bar{h})$ & 2& L (Light)\\
 $\mathcal{L}^-$ & $(J+1,\Delta-1)$ & $(h,1-\bar{h})$  & 2& F (Floodlight)\\
 $\mathcal{S} \equiv \mathcal{L}^+\mathcal{L}^-$ & $(2-\Delta,-J)$ & $(1-h,1-\bar{h})$  &2 & S (Full Shadow)\\ 
 $\mathcal{P}$ & $(\Delta,-J)$ & $(\bar{h},h)$ & 2 & S$_{J}$ (Spin-Shadow)\\ 
 $\mathcal{P}\mathcal{L}^+\mathcal{L}^-\equiv \mathcal{P}\mathcal{S}$ & $(2-\Delta,J)$ &  $(1-\bar{h},1-h)$  & 2 &S$_{\Delta}$ (Euclidean Shadow)\\ 
 $\mathcal{P}\mathcal{L}^+$ & $(1-J,\Delta-1)$ & $(\bar{h},1-h)$ & 4 & R\\
 $\mathcal{P}\mathcal{L}^-$ & $(J+1,1-\Delta)$ & $(1-\bar{h},h)$ & 4 & $\overline{\mbox{R}}$ \\
 \hline
\end{tabular}
\end{center}
\caption{We list the eight transformations that leave the quadratic Casimir invariant. The second column gives the spin and dimension of the transformed operators. The third column gives the conformal weights of the transformed operators. The fourth column gives the order of the operator. The last column compares the notation in this paper to that as defined in Table 1 of~\cite{2018JHEP...11..102K}.}
\label{tab}
\end{table}
Comparing to~\cite{2018JHEP...11..102K}, tells us that this ``parity" operator is the spin-shadow in 2D. In general, the spin-shadow transformation takes $J\rightarrow 2-d-J$ which is a non-trivial change in the spin. However, in $d=2$, it just flips the spin. In the context of celestial holography, $\mathcal{P}$ changes the helicity of a particle and does not give rise to an independent operator in the spectrum. Therefore, in our context\footnote{We would like to thank the referee from Physical Review D for pointing out this very crucial difference in $d=2$ compared to higher dimensions as it had significantly consequential effects on the results in this paper.} it does not make sense to talk about the four transformations that are compositions with $\mathcal{P}$.

It is, however, useful to note that these eight transformations form the dihedral group $D_8$, also denoted $D_4$ in geometry, which are the symmetries of a square. The elements of this group are written as
\begin{equation}
D_4 = \langle a,b: a^4=b^2=\mathbb{I},ab=ba^{-1}\rangle.
\end{equation}
Comparing to the above table, we can identify the two generators $a,b$ as $\mathcal{L}^+, \mathcal{P}$. All the other transformations can be written as compositions of these.\footnote{One could also have picked $\mathcal{L}^-, \mathcal{P}$ to be the generators.} 

\section{Yang Mills Marginal Operators}\label{sec:marginalops}

In order to obtain an operator that is exactly marginal using one of the above transformations, we need to start with an operator of spin $J=0,\pm 1$. While there are a multitude of reasons to consider scalars, for the sake of this work, we construct marginal operators starting from positive and negative helicity soft gluons. Starting with a positive helicity soft gluon $R^{+,a}$, we perform $\mathcal{L}^-$ to obtain a marginal operator. We will denote them as
\begin{equation}\label{eq:margpos}
\mathcal{M}_{\mathcal{L}^-}(z,\bar{z})  \equiv  -\frac{\alpha_+}{\sqrt{2}}T^a\mathcal{L}^-\left[R^{+,a}\right] = \frac{\alpha_+ T^a}{\sqrt{2\pi}}\int\frac{d\bar{w}}{(\bar{z}-\bar{w})^{2}}R^{+,a}(z,\bar{w}).
\end{equation}
Likewise, starting with a negative helicity soft gluon $R^{-,a}$ we perform $\mathcal{L}^+$ to obtain a marginal operator. We will denote them as 
\begin{equation}\label{eq:margneg}
\mathcal{M}_{\mathcal{L}^+}(z,\bar{z})  \equiv  \frac{\alpha_-}{\sqrt{2}}T^a\mathcal{L}^+\left[R^{-,a}\right] = \frac{\alpha_- T^a}{\sqrt{2\pi}} \int\frac{dw}{(z-w)^{2}}R^{-,a}(w,\bar{z}).
\end{equation}
The normalizations $\alpha_{\pm}$ in these definitions have been left arbitrary but should be chosen to appropriately normalize the two point functions. Scaling the operators arbitrarily will not affect the conclusions but will rather affect the actual explicit values of the quantities we will calculate, so our results are agnostic to these normalizations.

To develop a better intuition of what these operators actually are, it may be useful to find their explicit functional forms. To do this, one requires the conformal primary wavefunction for the $\Delta=1$ mode as in~\cite{Donnay:2018neh} which is what they refer to as the spin$-1$ Goldstone mode
\begin{equation}
A_{\mu;a}^G = \lim_{\Delta\rightarrow 1}A_{\mu;a}^{\Delta,\pm} = \partial_\mu \alpha_a^1, \ \ \alpha_a^1 = -\frac{\partial_a q\cdot X}{q\cdot X}
\end{equation}
where $a =\{w,\bar{w}\}$ denotes the helicity and $X$ is the four dimensional coordinate vector. Here $\partial_w q$ is the positive helicity polarization vector and is given by 
\begin{equation}
\partial_w\hat{q}^\mu(w,\bar{w}) = (\bar{w},1,-i,-\bar{w}).
\end{equation}
Given this, the relevant functional contribution to the light transform is (for positive helicity) is the light transform of the pure gauge parameter
\begin{equation}
\mathcal{L}^-(\alpha_z^1) = \int \frac{d\bar{w}}{(\bar{z}-\bar{w})^2}\frac{\partial_zq(z,\bar{w})\cdot X}{(-q(z,\bar{w})\cdot X)}.
\end{equation}
Since the numerator depends on $\bar{w}$, this works out slightly differently than what was computed in~\cite{Atanasov:2021cje}. We have that  
\begin{eqnarray}
\mathcal{L}^-(\alpha_z^1) & = & \int\frac{d\bar{w}}{(\bar{z}-\bar{w})^2}\frac{\partial_zq(z,\bar{w})\cdot X}{-q(z,\bar{w})\cdot X}\cr
& = & \int \frac{d\bar{w}}{(\bar{z}-\bar{w})^2}\frac{(X^1-X^2-\bar{w}(X^0+X^3))}{-q(z,\bar{w})\cdot X}\cr
& = & \frac{(\partial_zq(z,\bar{z})\cdot X)(\partial_{\bar{z}}q(z,\bar{z})\cdot X)}{(q(z,\bar{z})\cdot X)^2}\int d\bar{w}\bar{w}^{-2}(1-\bar{w})^{-1}\cr
& - & \frac{\partial_z\partial_{\bar{z}}q(z,\bar{z})\cdot X}{q(z,\bar{z})\cdot X}\int d\bar{w}\bar{w}^{-1}(1-\bar{w})^{-1}\cr
& = & \mathcal{N}_1\frac{\left(\partial_zq(z,\bar{z})\cdot X\right)\left(\partial_{\bar{z}}q(z,\bar{z})\cdot X\right)}{(q(z,\bar{z})\cdot X)^2} + \mathcal{N}_2\frac{\partial_z\partial_{\bar{z}}q(z,\bar{z})\cdot X}{q(z,\bar{z})\cdot X}
\end{eqnarray}
where the integrals are absorbed into normalization\footnote{These normalization factors are naively not finite and require some regularization to make them finite. This does not affect the results in the rest of the paper so we do not elaborate on this further here.} factors $\mathcal{N}_1,\mathcal{N}_2$. 

It is useful to check if this wavefunction has the appropriate scaling properties. If we scale $\bar{z}\rightarrow \lambda\bar{z}$ then $q\cdot X\rightarrow \sqrt{\lambda}q\cdot X$, $\partial_{\bar{z}}q\cdot X\rightarrow \frac{1}{\sqrt{\lambda}}\partial_{\bar{z}}q\cdot X$, $\partial_zq\cdot X\rightarrow \sqrt{\lambda}\partial_z q\cdot X$ and $\partial_z\partial_{\bar{z}}q\cdot X\rightarrow \frac{1}{\sqrt{\lambda}}\partial_z\partial_{\bar{z}}q\cdot X$ so this scales like $\lambda^{-1}$. Similarly, when $z\rightarrow\lambda z$ then $q\cdot X\rightarrow \sqrt{\lambda}q\cdot X$, $\partial_{\bar{z}}q\cdot X\rightarrow \sqrt{\lambda}\partial_{\bar{z}}q\cdot X$, $\partial_zq\cdot X\rightarrow \frac{1}{\sqrt{\lambda}}\partial_zq\cdot X$ and $\partial_z\partial_{\bar{z}}q\cdot X\rightarrow \frac{1}{\sqrt{\lambda}}\partial_z\partial_{\bar{z}}q\cdot X\rightarrow$ so it scales like $\lambda^{-1}$. Therefore this has weights $(h,\bar{h}) = (1,1)$ which means $\Delta=2, J=0$ as expected for a marginal operator. A similar expression exists for the other helicity. While this wavefunction contains some of the same ingredients, it's structure does not resemble that of any of the scalar conformal primary wavefunctions that we have seen. However, this may not be that surprising seeing that in~\cite{Atanasov:2021cje} upon light transforming a scalar, they did not get a wavefunction that had the same form. It should be noted that these were also computed in slightly different notation in~\cite{Sharma:2021gcz,Pasterski:2021fjn}.

Our goal is to extract any information that we can about the geometry of the space of CFTs. In~\cite{Kapec:2022axw} the authors looked at the four point function of marginal operators which encoded some non-singular data from the OPE, mainly the curvature of the manifold. However, for a given space the most basic data is the metric which is given by the two-point functions which we calculate first. Our starting point will be the color-ordered $\overline{\mbox{MHV}}$ amplitudes in 4D Yang-Mills. We will denote celestial $\overline{\mbox{MHV}}$ amplitudes as $\mathcal{A}^{a_1\cdots a_n}_{\Delta_1\cdots\Delta_n}(\vec{z}_1,\cdots,\vec{z}_n)$ where it is understood that the first two operators are the positive helicity operators and the rest are negative helicity. Correlators of marginal operators will be found by transforming the celestial amplitudes in accordance with~\eqref{eq:margpos} and~\eqref{eq:margneg}. It should be noted that one could have also considered the MHV amplitudes where there are two negative helicity operators and the rest positive. The calculations for those will proceed analogously and we will combine the results from both sets of amplitudes.

\section{Metric from Two Point Function}\label{sec:metric}
From~\eqref{eq:margOPE}, we see that the metric of the conformal manifold can be extracted from the two point function of marginal operators. Therefore, first we will look at the possible two point functions that we can construct using both of the light transforms. The celestial two point function of a positive and negative helicity gluon is~\cite{Sharma:2021gcz}
\begin{equation}
\mathcal{A}_{\Delta_1,\Delta_2}^{ab}(\vec{z}_1,\vec{z}_2) = 2\pi\delta(\Delta_1+\Delta_2-2)\delta^{ab}\delta(z_{12})\delta(\bar{z}_{12}).
\end{equation}
Naively, this two point function has a delta function singularity when both of the operators have dimension 1. In what follows, we will only be considering the light transformed two point function and the normalizations of the light transformed fields will account for this singularity.

The two point function of two marginal operators can be found by light transforming the above expression. Setting the overall normalization to 1, without loss of generality, we obtain
\begin{equation}
\langle \mathcal{M}_{\mathcal{L}^-}\mathcal{M}_{\mathcal{L}^+}\rangle  =  \frac{T^aT^b}{2\pi}\int \frac{d\bar{w}_1dw_2\mathcal{A}^{ab}_{1,1}(z_1,\bar{w}_1;w_2,\bar{z}_2)}{(\bar{z}_1-\bar{w}_1)^2(z_2-w_2)^2} = \frac{(N^2-1)}{2z_{12}^2\bar{z}_{12}^2}
\end{equation}
where we have used the convention for $SU(N)$ generators in the defining representation, that $T^aT^a = \frac{N^2-1}{2N}\mathbb{I}$ and $\mbox{Tr}\mathbb{I} = N$. Comparing to~\eqref{eq:margOPE} we can identify the metric on the conformal manifold as 
\begin{equation}
g_{IJ} = \frac{N^2-1}{2} \begin{bmatrix}0 & 1\cr 1 & 0\end{bmatrix}, \ \ \{I,J\} \in \{\mathcal{L}^-,\mathcal{L}^+\}.
\end{equation}
Again, the metric should be understood as $g_{IJ}(\lambda)|_{\lambda=0}$. The first aspect to notice is that the metric is inherently not diagonal. Given that the two point function pairs opposite helicity gluons, that is not surprising. It is, however, possible to diagonalize the metric and we see that the following linear combinations of operators will give rise to a diagonal metric
\begin{equation}
\mathcal{O}^\pm =\mathcal{M}_{\mathcal{L}^+}\pm\mathcal{M}_{\mathcal{L}^-}.
\end{equation}
We are able to make such a linear combination since both of the marginal operators are scalars. Therefore, the operators $\mathcal{O}^\pm$ are also scalars. Linear combinations of this type seem reminiscent of~\cite{Jorge-Diaz:2022dmy}, however in that paper they were considering combinations of in and out states so it is not exactly the same as what we have here. Regardless of the choice of linear combination we see that the conformal manifold is two dimensional. In the next section, we will discuss the three point functions which give rise to the leading order term in the OPE of marginal operators and whose coefficients correspond to the connections on this manifold. 

\section{Connection from Three-Point Function}\label{sec:connections}
The connections can be obtained from the leading term in the OPE which comes from the three point functions. The celestial three point function of two positive helicity gluons and one negative helicity gluon is~\cite{Sharma:2021gcz}
\begin{equation}
\mathcal{A}_{\Delta_1,\Delta_2,\Delta_3}^{abc}(\vec{z}_1,\vec{z}_2,\vec{z}_3) = 2i\pi f^{abc}\delta(\Delta_1+\Delta_2+\Delta_3-3)\Theta\left(\frac{\bar{z}_{13}}{\bar{z}_{12}}\right)\Theta\left(\frac{\bar{z}_{32}}{\bar{z}_{12}}\right)\frac{\delta(z_{13})\delta(z_{23})}{\bar{z}_{12}^{-\Delta_3}\bar{z}_{32}^{2-\Delta_1}\bar{z}_{13}^{2-\Delta_2}}.
\end{equation}
Note that this was obtained from the $\overline{\mbox{MHV}}$ amplitude but there is a similar expression for the MHV amplitude as well. The light transformed three point function
\begin{equation}
\langle \mathcal{M}_{\mathcal{L}^-}(z_1,\bar{z}_1)\mathcal{M}_{\mathcal{L}^-}(z_2,\bar{z}_2)\mathcal{M}_{\mathcal{L}^+}(z_3,\bar{z}_3)\rangle  = -\frac{N(N^2-1)}{4(2\pi)^{\frac{1}{2}}}\frac{\delta(z_{12})\delta(\bar{z}_{12})}{z_{32}^2\bar{z}_{32}^2}
\end{equation}
where we have used the convention $\mbox{Tr}(f^{abc}T^aT^bT^c) = \frac{iN}{4}(N^2-1)$. Pertinent details of this computation is contained in Appendix~\ref{app:correlators}. We observe that this three point function is distributional, which was expected given the properties of marginal operators we previously discussed. In particular it looks like $\delta^{(2)}(z_{12})$ which is consistent with the usual discussion in Euclidean CFTs~\cite{Kutasov:1988xb}.\footnote{We thank Sabrina Pasterski for pointing this out.}

One way to obtain the connections $\Gamma_{IJ}^K$ is to determine the leading term in the OPE by comparing the three point functions to the two point functions. This would be consistent with the method traditionally used to compute OPE coefficients in celestial CFT~\cite{Pate:2019lpp}. Another way we can find the connections is by integrating the possible three point functions as per~\eqref{eq:intthreept} and~\eqref{eq:gammas}. 

We will proceed the second way. Therefore, from the three point functions we obtain the following derivatives of the metric
\begin{equation}
 \frac{\partial g_{\mathcal{L}^-\mathcal{L}^+}}{\partial\lambda^{\mathcal{L}^-}} = \frac{\partial g_{\mathcal{L}^+\mathcal{L}^-}}{\partial\lambda^{\mathcal{L}^-}} =  -\frac{ N(N^2-1)}{4\sqrt{2\pi}}, \ \ \frac{\partial g_{\mathcal{L}^+\mathcal{L}^-}}{\partial\lambda^{\mathcal{L}^+}} = \frac{\partial g_{\mathcal{L}^-\mathcal{L}^+}}{\partial\lambda^{\mathcal{L}^+}} =  -\frac{ N(N^2-1)}{4\sqrt{2\pi}}
\end{equation}
where the latter two are found by extrapolating our results to the MHV sector. Recall that this definition required us to take the operators corresponding to the metric indices to positions $0,1$, which meant that if the delta function was in those two indices, it could never be satisfied and would yield zero. Lastly, it is important to note that since the metric is symmetric, i.e we can exchange the two operators in the two point function and it does not change, these derivatives are also symmetric in the exchange of the metric index labels which we have notated in the above equation. 

Substituting these into the expression for the connections will give us the following non-zero connections 
\begin{eqnarray}
\Gamma_{\mathcal{L}^+\mathcal{L}^+}^{\mathcal{L}^+} & = & \frac{1}{2}g^{\mathcal{L}^+\mathcal{L}^-}\left(\frac{\partial g_{\mathcal{L}^-\mathcal{L}^+}}{\partial\lambda^{\mathcal{L}^+}}+\frac{\partial g_{\mathcal{L}^-\mathcal{L}^+}}{\partial\lambda^{\mathcal{L}^+}}-\cancel{\frac{\partial g_{\mathcal{L}^+\mathcal{L}^+}}{\partial\lambda^{\mathcal{L}^-}}}\right) = -\frac{N(N^2-1)}{4\sqrt{2\pi}}\cr
\Gamma_{\mathcal{L}^-\mathcal{L}^-}^{\mathcal{L}^-} & = & \frac{1}{2}g^{\mathcal{L}^-\mathcal{L}^+}\left(\frac{\partial g_{\mathcal{L}^+\mathcal{L}^-}}{\partial\lambda^{\mathcal{L}^-}}+\frac{\partial g_{\mathcal{L}^+\mathcal{L}^-}}{\partial\lambda^{\mathcal{L}^-}}-\cancel{\frac{\partial g_{\mathcal{L}^-\mathcal{L}^-}}{\partial\lambda^{\mathcal{L}^+}}}\right) = -\frac{N(N^2-1)}{4\sqrt{2\pi}}
\end{eqnarray}
It is evident that we do not get any connections that mix opposite helicity operators. This is not surprising because the two-point function of same-helicity gluons vanishes but we will comment on the validity of this at the end of this section. Now that we have the connections and the metric components, we can write out the operator product expansion of the marginal operators without being concerned about ambiguities. This gives us the following OPEs of marginal operators
\begin{eqnarray}
\mathcal{M}_{\mathcal{L}^-}(z,\bar{z})\mathcal{M}_{J}(w,\bar{w}) & \sim & \frac{(N^2-1)\delta_{J\mathcal{L}^+}}{2|z-w|^4} -\frac{N(N^2-1)}{4\sqrt{2\pi}}\delta_{J\mathcal{L}^-}\mathcal{M}_{\mathcal{L}^-}(w,\bar{w})\delta^{(2)}(z-w)\cr
\mathcal{M}_{\mathcal{L}^+}(z,\bar{z})\mathcal{M}_{J}(w,\bar{w})  & \sim &  \frac{(N^2-1)\delta_{J\mathcal{L}^-}}{2|z-w|^4} -\frac{N(N^2-1)}{4\sqrt{2\pi}}\delta_{J\mathcal{L}^+}\mathcal{M}_{\mathcal{L}^+}(w,\bar{w})\delta^{(2)}(z-w)
\end{eqnarray}
As we mentioned before, we have specifically chosen not to work in local coordinates where the connections vanish. If we had, all the distributional terms in the above OPEs would vanish. While that would be perfectly fine for understanding the OPEs of the marginal operators, the presence of the distributional terms above are important for extrapolating the operators appearing in the soft OPEs. Since the marginal operators are transformed soft gluons, in principle one can backtrack and deduce what soft operators we expect to find in their OPEs. Unsurprisingly what we find is that these terms are consistent with the OPEs that we already have of soft gluons\footnote{We don't usually write down the OPEs of mixed helicity soft gluons since we usually care about one chiral sector of the soft algebra.} because they say that $++\rightarrow +$ and $--\rightarrow -$. 

It is also important to note here that the distributional terms are consistent\footnote{Up to a factor of $N^2-1$ which comes from the fact that we did not normalize the two point function's coefficient to be 1.} with what we would have gotten by comparing our two and three point functions of light transformed operators. However, there is the possibility of missed OPE terms due to the vanishing of a certain set of three point functions. Namely, connections that result in mixed helicity terms in the OPE are absent since the three point functions that they would originate from vanish in the MHV sector. This is generally an issue when extracting OPE coefficients by comparing three point and two point MHV amplitudes. If we considered higher order correlators then we would most likely be able to extract these sorts of terms. Therefore, this method of extracting OPE data does not appear to be any better than what has already been done but it is nonetheless nice to see it is comparable.

\section{Discussion}\label{sec:discussion}
In~\cite{Kapec:2022axw} the authors focused on extracting the curvature of the manifold from the four point function of marginal operators. However, the most fundamental description of the geometry comes from the metric which one can deduce from the two point function. In the case of Yang-Mills, we find that there exists a set of marginal operators giving us two independent directions that we can travel on the conformal manifold. While we are fairly certain that there are no more than two directions, since that would imply being able to construct some new set of marginal operators, it is not clear if the above analysis gives us the full metric. The components of the metric we determined were based solely on the existence of non-zero two point functions. While it is true that the only non-vanishing two point function is between opposite helicity gluons, it is not clear whether that necessarily implies the non-existence of a two point function between $\mathcal{L}^-$ and $\mathcal{L}^-$, for instance. 

In fact, starting from the three point functions one could ask if the metric derivatives we wrote down are the only ones that exist. Since the definition of $\frac{\partial g_{IJ}}{\partial\lambda^K}$, takes the three point function where $\mathcal{M}_K$ is a function of the variables $(z,\bar{z})$ we integrate over, if we take the same three point function and set a different operator to have the argument $(z,\bar{z})$, that will lead to a different combination of indices like $\frac{\partial g_{IK}}{\partial\lambda^J}$, for instance. However, if we look at the three point functions they all contain $\delta(\bar{z}_{12})$, which means that if we set the argument of the third operator to be the one integrated over, the delta function will be $\delta(1)$ which is 0. Therefore, given this set of non-vanishing three point functions, we have derived the only non-zero metric derivatives that exist. It would be interesting to see what happens if one takes the full OPE of soft gluons (including mixed helicity terms) and works backwards to find the metric and connections. This could give rise to other two point and three-point functions of marginal operators that are not enumerated here.

It is additionally interesting to speculate on the connections to the statements made in~\cite{Cheung:2022vnd} where they mention how to understand the geometric soft theorems in gauge theory and gravity by relating them to a sigma model. The authors propose a way to tack on the necessary kinematic structure to the nonlinear sigma model field that is consistent with gauge theory and gravity. While the authors have expressions for the metric, connections and curvature it is not clear how those explicitly relate to what we have found in this work. There are small ingredients that are similar, like the existence of $\delta^{ab}$ in the metric and $f^{abc}$ in the connection, but those could just be a result of working with quantities in a theory of Yang-Mills. Additionally, the authors of~\cite{Cheung:2022vnd} work specifically with Yang-Mills in Feynman gauge and the expressions they propose for the metric and connections are gauge dependent while in this paper we have made statements that are all agnostic to the gauge. It would be incredibly useful if we could completely decouple from the bulk theory but at the present moment, that seems rather difficult.

Simultaneously, it is useful to point out a subtlety when comparing this work to that done in~\cite{Kapec:2022axw,Kapec:2022hih}. In both of the previous papers, there was a known set of vacua in the bulk spacetime so that moving along the conformal manifold was understood as moving from one vacuum state to another. In~\cite{Kapec:2022axw} since the un-shadowed operators were scalars, the shadowed operators were exactly marginal, the boundary symmetries were preserved and marginal deformations corresponded to moving around in the space of vacua of the bulk sigma model. In~\cite{Kapec:2022hih} the operators were not marginal in the traditional sense and deforming by these operators actually shifted from one copy of the boundary conformal group to another. Therefore, strictly speaking, the conformal symmetry was broken but then restored which allowed for a deformation of this type that was not quite marginal.\footnote{We thank Daniel Kapec for highlighting this point.} In this work, the operators are marginal but do not have an interpretation as moving around in the space of vacua corresponding to large gauge transformations. However, a discussion of this type involving the geometry of a conformal manifold suggests that there is such an interpretation. At this time, we do not understand what space of vacua this could describe but it is definitely a very interesting question left for future work.

Lastly, we should discuss the case of gravity since that is what we endeavor to understand. Ideally, we would like to understand marginal operators in the case where the bulk theory was a theory of quantum gravity, however it seems like if you begin with an operator of spin $J=\pm 2$, then its dimension must be complex in order to transform it to a marginal operator. Therefore, this particular logic does not work for the graviton case but it is possible that there is some way to shift the spin or dimension, maybe via the use of a descendent rather than a primary or by considering composite operators, in order to discuss marginal operators in that case. We also leave this for future exploration.

\section*{Acknowledgements}
I am grateful to Erin Crawley, Dan Kapec, Walker Melton, Rajamani Narayanan, Sabrina Pasterski and Ana-Maria Raclariu for useful discussions. Additional thanks to Sabrina Pasterski and Dan Kapec for looking at a preliminary draft. I would also like to thank the referee from Physical Review D for their useful comments and, in particular, for catching a flaw in the previous version of this paper that I, regretfully, missed. The author acknowledges support by the Celestial Holography Initiative at the Perimeter Institute for Theoretical Physics and by the Simons Collaboration on Celestial Holography. The author's research at the Perimeter Institute is supported by the Government of Canada through the Department of Innovation, Science and Industry Canada and by the Province of Ontario through the Ministry of Colleges and Universities.

\appendix
\section{Action of Composite Integral Transforms}\label{app:integrals}
In this appendix we will compute the action of a composite integral transforms on an operator in the event that this is every useful for someone. We will consider arbitrary transformations so that we can use this to solve for the forms of explicit transformations. Let us take integral transforms of the following form
\begin{equation}
\mathcal{T}_i\left[\mathcal{O}\right](w,\bar{w}) = \mathcal{C}^i_{h,\bar{h}}\int \frac{d^2 z}{(w-z)^{f_i(h,\bar{h})}(\bar{w}-\bar{z})^{g_i(h,\bar{h})}}\mathcal{O}_{h,\bar{h}}(z,\bar{z})
\end{equation}
where the initial operator has weights $(h,\bar{h})$ and the transformed operator has weights $(h_i,\bar{h}_i)$. The action of a composite operator is 
\begin{equation}
\mathcal{T}_i\mathcal{T}_j\left[\mathcal{O}\right](w,\bar{w}) = \mathcal{C}^i_{h_j,\bar{h}_j}\mathcal{C}^j_{h_,\bar{h}}\int \frac{d^2z_2d^2z_1\mathcal{O}_{h,\bar{h}}(z_1,\bar{z}_1)}{(w-z_2)^{f_i(h_j,\bar{h}_j)}(\bar{w}-\bar{z}_2)^{g_i(h_j,\bar{h}_j)}(z_2-z_1)^{f_j(h,\bar{h})}(\bar{z}_2-\bar{z}_1)^{g_j(h,\bar{h})}}.
\end{equation}
In order to determine whether this composite transformation is equivalent to another transformation, we need to perform the intermediary integral which is the one over $z_2,\bar{z}_2$. We can first rewrite it as 
\begin{eqnarray}
\mathcal{I} & = & \int \frac{d^2z_2(-1)^{f_j(h,\bar{h})+g_j(h,\bar{h})}}{(w-z_2)^{f_i(h_j,\bar{h}_j)}(\bar{w}-\bar{z}_2)^{g_i(h_j,\bar{h}_j)}(z_1-z_2)^{f_j(h,\bar{h})}(\bar{z}_1-\bar{z}_2)^{g_j(h,\bar{h})}}\cr
& = & (-1)^{f_j(h,\bar{h})+g_j(h,\bar{h})}\int \frac{d^2z_2 (w-z_2)^{g_i(h_j,\bar{h}_j)-f_i(h_j,\bar{h}_j)}(z_1-z_2)^{g_j(h,\bar{h})-f_j(h,\bar{h})}}{|w-z_2|^{2g_i(h_j,\bar{h}_j)}|z_1-z_2|^{2g_j(h,\bar{h})}}.
\end{eqnarray}
Then we can inverse Fourier transform twice to put it in the following form
\begin{eqnarray}
\mathcal{I} & = & (-1)^{f_j(h,\bar{h})+g_j(h,\bar{h})}\int d^2 z_2\int\frac{d^2k_1}{(2\pi)^2}\frac{d^2k_2}{(2\pi)^2}\int d^2x_1 d^2 x_2 \frac{x_1^{g_i(h_j,\bar{h}_j)-f_i(h_j,\bar{h}_j)}x_2^{g_j(h,\bar{h})-f_j(h,\bar{h})}}{(x_1\bar{x}_1)^{g_i(h_j,\bar{h}_j)}(x_2\bar{x}_2)^{g_j(h,\bar{h})}}\cr
&\times & \exp\left[ik_1\cdot(x_1-w+z_2)+ik_2\cdot(x_2-z_1+z_2)\right].
\end{eqnarray}
The integral over $z_2,\bar{z}_2$ yields a delta function which can be used to do one of the momentum integrals to give
\begin{eqnarray}
\mathcal{I} & = & 2(-1)^{f_j(h,\bar{h})+g_j(h,\bar{h})}\int\frac{d^2k_1}{(2\pi)^2}e^{-ik_1\cdot(w-z_1)}\int d^2x_1 d^2 x_2 \frac{x_1^{g_i(h_j,\bar{h}_j)-f_i(h_j,\bar{h}_j)}x_2^{g_j(h,\bar{h})-f_j(h,\bar{h})}e^{ik_1\cdot x_{12}}}{(x_1\bar{x}_1)^{g_i(h_j,\bar{h}_j)}(x_2\bar{x}_2)^{g_j(h,\bar{h})}}.
\end{eqnarray}
Next we can use Schwinger parametrization to rewrite the denominator and then perform the $x_i$ integrals
\begin{eqnarray}
\mathcal{I} & = & \frac{2(-1)^{f_j(h,\bar{h})+g_j(h,\bar{h})}}{\Gamma(g_i(h_j,\bar{h}_j))\Gamma(g_j(h,\bar{h}))}\int\frac{d^2k_1}{(2\pi)^2}e^{-ik_1\cdot(w-z_1)}\int_0^\infty d\beta_1 d\beta_2 \beta_1^{g_i(h_j,\bar{h}_j)-1}\beta_2^{g_j(h,\bar{h})-1} \cr
& \times & \int d^2x_1 d^2 x_2x_1^{g_i(h_j,\bar{h}_j)-f_i(h_j,\bar{h}_j)}x_2^{g_j(h,\bar{h})-f_j(h,\bar{h})}e^{-\beta_1x_1\bar{x}_1-\beta_2x_2\bar{x}_2+ik_1\cdot(x_1-x_2)}\cr
& = & \frac{2\pi^2i^{g_i(h_j,\bar{h}_j)-f_i(h_j,\bar{h}_j)+g_j(h,\bar{h})-f_j(h,\bar{h})}}{\Gamma(g_i(h_j,\bar{h}_j))\Gamma(g_j(h,\bar{h}))}\int\frac{d^2k_1}{(2\pi)^2}e^{-ik_1\cdot(w-z_1)}\cr
& \times &\bar{k}_1^{g_i(h_j,\bar{h}_j)-f_i(h_j,\bar{h}_j)+g_j(h,\bar{h})-f_j(h,\bar{h})}\int_0^\infty d\beta_1 d\beta_2 \beta_1^{f_i(h_j,\bar{h}_j)-2}\beta_2^{f_j(h,\bar{h})-2}e^{-\frac{k_1\bar{k}_1}{\beta_1}-\frac{k_1\bar{k}_1}{\beta_2}}\cr
& = & \frac{\pi^2\Gamma(1-f_i(h_j,\bar{h}_j))\Gamma(1-f_j(h,\bar{h}))i^{g_i(h_j,\bar{h}_j)-f_i(h_j,\bar{h}_j)+g_j(h,\bar{h})-f_j(h,\bar{h})}}{2\Gamma(g_i(h_j,\bar{h}_j))\Gamma(g_j(h,\bar{h}))}\cr
& \times &\int\frac{d^2k_1}{(2\pi)^2}e^{-ik_1\cdot(w-z_1)}\frac{k_1^{f_i(h_j,\bar{h}_j)+f_j(h,\bar{h})-g_i(h_j,\bar{h}_j)-g_j(h,\bar{h})}}{(k_1\bar{k}_1)^{2-g_i(h_j,\bar{h}_j)-g_j(h,\bar{h})}}\cr
& = & \frac{\pi(-1)^{f_i(h_j,\bar{h}_j)+f_j(h,\bar{h})-g_i(h_j,\bar{h}_j)-g_j(h,\bar{h})}\Gamma(1-f_i(h_j,\bar{h}_j))\Gamma(1-f_j(h,\bar{h}))}{\Gamma(g_i(h_j,\bar{h}_j))\Gamma(g_j(h,\bar{h}))\Gamma(2-g_i(h_j,\bar{h}_j)-g_j(h,\bar{h}))}\cr
& \times & \int_0^\infty d\beta \beta^{-f_i(h_j,\bar{h}_j)-f_j(h,\bar{h})}(\bar{w}-\bar{z}_1)^{f_i(h_j,\bar{h}_j)+f_j(h,\bar{h})-g_i(h_j,\bar{h}_j)-g_j(h,\bar{h})}e^{-\frac{|w-z_1|^2}{\beta}}\cr
& = & \frac{\pi(-1)^{f_i(h_j,\bar{h}_j)+f_j(h,\bar{h})-g_i(h_j,\bar{h}_j)-g_j(h,\bar{h})}\Gamma(1-f_i(h_j,\bar{h}_j))\Gamma(1-f_j(h,\bar{h}))}{\Gamma(g_i(h_j,\bar{h}_j))\Gamma(g_j(h,\bar{h}))}\cr
& \times &\frac{\Gamma(f_i(h_j,\bar{h}_j)+f_j(h,\bar{h})-1)}{\Gamma(2-g_i(h_j,\bar{h}_j)-g_j(h,\bar{h}))} (w-z_1)^{1-f_i(h_j,\bar{h}_j)-f_j(h,\bar{h})}(\bar{w}-\bar{z}_1)^{1-g_i(h_j,\bar{h}_j)-g_j(h,\bar{h})}
\end{eqnarray}
One fixes the normalization of these operators depending on whether the composite operator is the shadow or the identity. If the product of two operators gives the identity then the functions must satisfy
\begin{equation}
f_i(h_j,\bar{h}_j)+f_j(h,\bar{h}) = g_i(h_j,\bar{h}_j) + g_j(h,\bar{h}) = 2.
\end{equation}
Then we have~\footnote{From~\cite{Dolan:2011dv} there are actually two possible choices for this integral product but we have chosen the one that allows us to fulfill the other constraints simultaneously.}
\begin{eqnarray}
\mathcal{T}_i\mathcal{T}_j\left[\mathcal{O}\right](w,\bar{w}) &= & \mathcal{C}^i_{h_j,\bar{h}_j}\mathcal{C}^j_{h_,\bar{h}}\int \frac{d^2z_2d^2z_1\mathcal{O}_{h,\bar{h}}(z_1,\bar{z}_1)}{(w-z_2)^{f_i(h_j,\bar{h}_j)}(\bar{w}-\bar{z}_2)^{g_i(h_j,\bar{h}_j)}(z_2-z_1)^{f_j(h,\bar{h})}(\bar{z}_2-\bar{z}_1)^{g_j(h,\bar{h})}}\cr
& = & \mathcal{C}^i_{h_j,\bar{h}_j}\mathcal{C}^j_{h_,\bar{h}}\frac{\pi^2 \Gamma(1-g_i(h_j,\bar{h}_j))\Gamma(1-g_j(h,\bar{h})}{\Gamma(f_i(h_j,\bar{h}_j))\Gamma(f_j(h,\bar{h}))}\int d^2z_1\delta^{(2)}(w-z_1)\mathcal{O}_{h,\bar{h}}(z_1,\bar{z}_1)\cr
& = & \mathcal{C}^i_{h_j,\bar{h}_j}\mathcal{C}^j_{h_,\bar{h}}\frac{\pi^2 \Gamma(1-g_i(h_j,\bar{h}_j))\Gamma(1-g_j(h,\bar{h})}{\Gamma(f_i(h_j,\bar{h}_j))\Gamma(f_j(h,\bar{h}))}\mathcal{O}_{h,\bar{h}}(w,\bar{w}).
\end{eqnarray}
This implies that the normalizations should satisfy the following equation
\begin{equation}
\mathcal{C}^i_{h_j,\bar{h}_j}\mathcal{C}^j_{h,\bar{h}}  = \frac{\Gamma(f_i(h_j,\bar{h}_j))\Gamma(f_j(h,\bar{h}))}{\pi^2\Gamma(1-g_i(h_j,\bar{h}_j))\Gamma(1-g_j(h,\bar{h}))}.
\end{equation}
It is easy to see that in the case of the identity the normalizations are $\mathcal{C}^i_{h,\bar{h}} = \frac{\Gamma(f_i(h,\bar{h}))}{\pi\Gamma(1-g_i(h,\bar{h}))}$.  If the product of two operators is the shadow, the powers have to satisfy
\begin{eqnarray}\label{eq:shadowpowers}
f_i(h_j,\bar{h}_j)+f_j(h,\bar{h})-1 & = & 2-2h\cr
g_i(h_j,\bar{h}_j)+g_j(h,\bar{h})-1 & = & 2-2\bar{h}.
\end{eqnarray}
Therefore, we know that the normalizations should satisfy the following equation
\begin{equation}
\mathcal{C}^i_{h_j,\bar{h}_j}\mathcal{C}^j_{h,\bar{h}}\frac{\pi\Gamma(1-f_i(h_j,\bar{h}_j))\Gamma(1-f_j(h,\bar{h}))\Gamma(2-2h)}{\Gamma(g_i(h_j,\bar{h}_j))\Gamma(g_j(h,\bar{h}))\Gamma(2\bar{h}-1)} = \frac{\Gamma(2-2h)}{\pi\Gamma(2\bar{h}-1)}.
\end{equation}
which we can see is satisfied using~\eqref{eq:shadowpowers}.

\section{Calculation of Correlators}\label{app:correlators}
For the three point functions we will need this general integral
\begin{eqnarray}
\mathcal{I}_{\alpha\beta\gamma} & = & \int\frac{d\bar{w}_1}{(\bar{z}_1-\bar{w}_1)^\alpha}\frac{d\bar{w}_2}{(\bar{z}_2-\bar{w}_2)^\beta}\frac{dw_3}{(z_3-w_3)^\gamma}\Theta\left(\frac{\bar{w}_{13}}{\bar{w}_{12}}\right)\Theta\left(\frac{\bar{w}_{32}}{\bar{w}_{12}}\right)\frac{\delta(w_{13})\delta(w_{23})}{\bar{w}_{12}^{-\Delta_3}\bar{w}_{32}^{2-\Delta_1}\bar{w}_{13}^{2-\Delta_2}}\cr
& = & \frac{\delta(w_{21})}{(z_3-w_1)^\gamma}\int\frac{d\bar{w}_1}{(\bar{z}_1-\bar{w}_1)^\alpha}\frac{d\bar{w}_2}{(\bar{z}_2-\bar{w}_2)^\beta}\frac{\Theta\left(\bar{w}_{1}-\bar{w}_3\right)\Theta\left(\bar{w}_{3}-\bar{w}_2\right)}{(\bar{w}_{1}-\bar{w}_2)^{-\Delta_3}(\bar{w}_{3}-\bar{w}_2)^{2-\Delta_1}(\bar{w}_{1}-\bar{w}_3)^{2-\Delta_2}}\cr
& = & \frac{(-1)^{\alpha}\delta(w_{21})}{(z_3-w_1)^\gamma}\int\frac{d\bar{w}_1}{(\bar{w}_1+\bar{w}_3-\bar{z}_1)^\alpha}\frac{d\bar{w}_2}{(\bar{w}_2-\bar{w}_3+\bar{z}_2)^\beta}\frac{\Theta\left(\bar{w}_{1}\right)\Theta\left(\bar{w}_2\right)}{(\bar{w}_{1}+\bar{w}_2)^{-\Delta_3}\bar{w}_2^{2-\Delta_1}\bar{w}_{1}^{2-\Delta_2}}\cr
& = & \frac{(-1)^{\alpha}\delta(w_{21})}{(z_3-w_1)^\gamma}\sum_{k=-\infty}^\infty \frac{\Gamma(1+\Delta_3)}{\Gamma(k+1)\Gamma(1+\Delta_3-k)}\int_0^\infty\frac{d\bar{w}_1\bar{w}_1^{k-2+\Delta_2}}{(\bar{w}_1+\bar{w}_3-\bar{z}_1)^\alpha}\int_0^\infty\frac{d\bar{w}_2\bar{w}_2^{\Delta_3-k-2+\Delta_1}}{(\bar{w}_2-\bar{w}_3+\bar{z}_2)^\beta}\cr
& = & \frac{(-1)^{\alpha}\delta(w_{21})}{(z_3-w_1)^\gamma}\sum_{k=-\infty}^\infty \frac{\Gamma(1+\Delta_3)}{\Gamma(k+1)\Gamma(1+\Delta_3-k)}\cr
& \times & \frac{(\bar{w}_3-\bar{z}_1)^{k-1+\Delta_2-\alpha}\Gamma(k-1+\Delta_2)\Gamma(-k+1-\Delta_2+\alpha)}{\Gamma(\alpha)}\cr
&\times & \frac{(\bar{z}_2-\bar{w}_3)^{\Delta_3-k-1+\Delta_1-\beta}\Gamma(\Delta_3-k-1+\Delta_1)\Gamma(1-\Delta_3+k-\Delta_1+\beta)}{\Gamma(\beta)}.
\end{eqnarray}
We note that we have implicitly assumed $\bar{w}_1>\bar{w}_3>\bar{w}_2$ but one gets the same answer assuming that $\bar{w}_2>\bar{w}_3>\bar{w}_1$. Therefore this is the answer regardless of which regime we choose given the Heaviside function constraints. Where we have used the generalized binomial theorem to perform the integrals. We only care about this when $\Delta_1=\Delta_2=\Delta_3=1$ and $\alpha=\beta=\gamma=2$
\begin{eqnarray}
\mathcal{I}_{2,2,2}(w_1,\bar{z}_1;w_2,\bar{z}_2;z_3,\bar{w}_3) & = & \frac{\delta(w_{21})\delta(\bar{z}_{21})}{(z_3-w_1)^2(\bar{w}_3-\bar{z}_1)^{2}}.
\end{eqnarray}
The relevant three point function is then
\begin{equation}
\langle \mathcal{M}_{\mathcal{L}^-}\mathcal{M}_{\mathcal{L}^-}\mathcal{M}_{\mathcal{L}^+}\rangle  =  \frac{if^{abc}T^aT^bT^c}{(2\pi)^{\frac{1}{2}}}\mathcal{I}_{2,2,2}(z_1,\bar{z}_1;z_2,\bar{z}_2;z_3,\bar{z}_3)=  \frac{if^{abc}T^aT^bT^c}{(2\pi)^{\frac{1}{2}}}\frac{\delta(z_{12})\delta(\bar{z}_{12})}{z_{31}^2\bar{z}_{31}^2}.
\end{equation}
One can also check this formula by transforming the three point function. First we calculate the stripped three point function, which does not have the normalizations of the light transform
\begin{eqnarray}
\mathcal{L}_3^{++-}(\vec{z}_1,\vec{z}_2,\vec{z}_3) & = & 2\pi\delta(\Delta_1+\Delta_2+\Delta_3-3)\int\frac{d\bar{w}_1}{(\bar{z}_1-\bar{w}_1)^{3-\Delta_1}}\frac{d\bar{w}_2}{(\bar{z}_2-\bar{w}_2)^{3-\Delta_2}}\frac{dw_3}{(z_3-w_3)^{3-\Delta_3}}\cr
& \times & \Theta\left(\frac{\bar{w}_1-\bar{z}_3}{\bar{w}_1-\bar{w}_2}\right)\Theta\left(\frac{\bar{z}_3-\bar{w}_2}{\bar{w}_1-\bar{w}_2}\right)\frac{\delta(z_1-w_3)\delta(z_2-w_3)}{\bar{w}_{12}^{-\Delta_3}(\bar{z}_3-\bar{w}_2)^{2-\Delta_1}(\bar{w}_1-\bar{z}_3)^{2-\Delta_2}}\cr
& = & \frac{2\pi\delta(\Delta_1+\Delta_2+\Delta_3-3)\delta(z_{12})}{z_{32}^{3-\Delta_3}}\int\frac{d\bar{w}_1}{(\bar{z}_1-\bar{w}_1)^{3-\Delta_1}}\frac{d\bar{w}_2}{(\bar{z}_2-\bar{w}_2)^{3-\Delta_2}}\cr
& \times & \Theta\left(\frac{\bar{w}_1-\bar{z}_3}{\bar{w}_1-\bar{w}_2}\right)\Theta\left(\frac{\bar{z}_3-\bar{w}_2}{\bar{w}_1-\bar{w}_2}\right)\frac{1}{\bar{w}_{12}^{-\Delta_3}(\bar{z}_3-\bar{w}_2)^{2-\Delta_1}(\bar{w}_1-\bar{z}_3)^{2-\Delta_2}}
\end{eqnarray}
If $\bar{w}_1>\bar{w}_2$, then $\bar{w}_1>\bar{z}_3$ and $\bar{w}_2<\bar{z}_3$
\begin{eqnarray}
\mathcal{L}_3^{++-}(\vec{z}_1,\vec{z}_2,\vec{z}_3) & = &\frac{2\pi\delta(\Delta_1+\Delta_2+\Delta_3-3)\delta(z_{12})}{z_{32}^{3-\Delta_3}}\int\frac{d\bar{w}_1}{(\bar{z}_1-\bar{w}_1)^{3-\Delta_1}}\int \frac{d\bar{w}_2}{(\bar{z}_2-\bar{w}_2)^{3-\Delta_2}}\cr
& \times & \frac{1}{\bar{w}_{12}^{-\Delta_3}(\bar{z}_3-\bar{w}_2)^{2-\Delta_1}(\bar{w}_1-\bar{z}_3)^{2-\Delta_2}}\Theta(\bar{w}_1-\bar{z}_3)\Theta(\bar{z}_3-\bar{w}_2)\cr
& = & \frac{2\pi\delta(\Delta_1+\Delta_2+\Delta_3-3)\delta(z_{12})}{z_{32}^{3-\Delta_3}}\sum_{k=-\infty}^\infty \frac{\Gamma(\Delta_3+1)(-1)^{\Delta_1-3}}{\Gamma(\Delta_3-k+1)\Gamma(k+1)}\cr
&\times & \int_0^\infty d\bar{w}_1 \bar{w}_1^{\Delta_3+\Delta_2-2-k}(\bar{w}_1-\bar{z}_{13})^{\Delta_1-3}\int_0^\infty d\bar{w}_2\bar{w}_2^{\Delta_1+k-2}(\bar{z}_{23}+\bar{w}_2)^{\Delta_2-3}\cr
& = & \frac{2\pi\delta(\Delta_1+\Delta_2+\Delta_3-3)\delta(z_{12})}{z_{32}^{3-\Delta_3}}\sum_{k=-\infty}^\infty \frac{\Gamma(\Delta_3+1)(-1)^{\Delta_1-3-1-k}}{\Gamma(\Delta_3-k+1)\Gamma(k+1)}\cr
&\times & \bar{z}_{23}^{\Delta_1+k+\Delta_2-4}\bar{z}_{13}^{-1-k}B(k+1,2-\Delta_1-k) B(1+\Delta_3-k,\Delta_1+k-1)\cr
& = & \frac{2\pi(-1)^{-\Delta_2}\delta(\Delta_1+\Delta_2+\Delta_3-3)\Gamma(\Delta_3+1)\delta(z_{12})}{\Gamma(\Delta_3+\Delta_1)\Gamma(3-\Delta_1)z_{32}^{3-\Delta_3}\bar{z}_{32}^{1+\Delta_3}\bar{z}_{13}} \cr
&\times & \sum_{k=-\infty}^\infty\left(\frac{\bar{z}_{32}}{\bar{z}_{13}}\right)^{k}\Gamma(2-\Delta_1-k) \Gamma(\Delta_1+k-1)\cr
& = & \frac{2\pi(-1)^{1-\Delta_2}\delta(\Delta_1+\Delta_2+\Delta_3-3)\Gamma(\Delta_3+1)\Gamma(\Delta_1)\Gamma(1-\Delta_1)\delta^2(z_{12})}{\Gamma(3-\Delta_2)\Gamma(3-\Delta_1)z_{32}^{3-\Delta_3}\bar{z}_{32}^{1+\Delta_3}} 
\end{eqnarray}
In the case where $\bar{w}_1<\bar{w}_2$, the signs in the theta functions have to change and we have 
\begin{eqnarray}
\mathcal{L}_3^{++-}(\vec{z}_1,\vec{z}_2,\vec{z}_3) & = &\frac{2\pi\delta(\Delta_1+\Delta_2+\Delta_3-3)\delta(z_{12})}{z_{32}^{3-\Delta_3}}\int\frac{d\bar{w}_1}{(\bar{z}_1-\bar{w}_1)^{3-\Delta_1}}\int \frac{d\bar{w}_2}{(\bar{z}_2-\bar{w}_2)^{3-\Delta_2}}\cr
& \times & \frac{1}{\bar{w}_{12}^{-\Delta_3}(\bar{z}_3-\bar{w}_2)^{2-\Delta_1}(\bar{w}_1-\bar{z}_3)^{2-\Delta_2}}\Theta(\bar{z}_3-\bar{w}_1)\Theta(\bar{w}_2-\bar{z}_3)\cr
& = & -\frac{2\pi\delta(\Delta_1+\Delta_2+\Delta_3-3)\delta(z_{12})}{z_{32}^{3-\Delta_3}}\sum_{k=-\infty}^\infty \frac{\Gamma(\Delta_3+1)(-1)^{3-\Delta_2}}{\Gamma(\Delta_3-k+1)\Gamma(k+1)}\cr
&\times & \int_0^\infty d\bar{w}_1(\bar{w}_1+\bar{z}_{13})^{\Delta_1-3}\bar{w}_1^{1-\Delta_1-k}\int_0^\infty d\bar{w}_2(\bar{w}_2-\bar{z}_{23})^{\Delta_2-3}\bar{w}_2^{\Delta_1+k-2}\cr
& = & -\frac{2\pi\delta(\Delta_1+\Delta_2+\Delta_3-3)\delta(z_{12})}{z_{32}^{3-\Delta_3}}\sum_{k=-\infty}^\infty \frac{\Gamma(\Delta_3+1)(-1)^{3-\Delta_2}}{\Gamma(\Delta_3-k+1)\Gamma(k+1)}\cr
&\times & \bar{z}_{32}^{-\Delta_3-1+k}\bar{z}_{13}^{-k-1}B(1+k,2-\Delta_1-k) B(1+\Delta_3-k,\Delta_1+k-1)\cr
& = & -\frac{2\pi\delta(\Delta_1+\Delta_2+\Delta_3-3)\Gamma(\Delta_3+1)(-1)^{3-\Delta_2}\delta(z_{12})}{\Gamma(3-\Delta_1)\Gamma(\Delta_1+\Delta_3)z_{32}^{3-\Delta_3}\bar{z}_{32}^{1+\Delta_3}\bar{z}_{13}} \cr
&\times & \sum_{k=-\infty}^\infty\left(\frac{\bar{z}_{32}}{\bar{z}_{13}}\right)^{k}\Gamma(2-\Delta_1-k)\Gamma(\Delta_1+k-1)\cr
& = & \frac{2\pi(-1)^{1-\Delta_2}\delta(\Delta_1+\Delta_2+\Delta_3-3)\Gamma(\Delta_3+1)\Gamma(1-\Delta_1)\Gamma(\Delta_1) \delta^2(z_{12})}{\Gamma(\Delta_1+\Delta_3)\Gamma(3-\Delta_1)z_{32}^{3-\Delta_3}\bar{z}_{32}^{1+\Delta_3}}. 
\end{eqnarray}
We will be concerned with this three point function for the case where $\Delta_1=\Delta_2=\Delta_3=1$. This gives us 
\begin{eqnarray}
\mathcal{L}_3^{++-}(\vec{z}_1,\vec{z}_2,\vec{z}_3) & = &\frac{2\pi \delta^2(z_{12})}{z_{32}^{2}\bar{z}_{32}^{2}}
\end{eqnarray}
which is in agreement up to prefactors.

\bibliographystyle{JHEP}
 \bibliography{marginalgluons.bib}

\providecommand{\href}[2]{#2}\begingroup\raggedright\begin{thebibliography}{10}

\bibitem{DiFrancesco:639405}
P.~Di~Francesco, P.~Mathieu and D.~Sénéchal, \emph{{Conformal field theory}},
  Graduate texts in contemporary physics. Springer, New York, NY, 1997,
  \href{https://doi.org/10.1007/978-1-4612-2256-9}{10.1007/978-1-4612-2256-9}.

\bibitem{Raclariu:2021zjz}
A.-M. Raclariu, \emph{{Lectures on Celestial Holography}},
  \href{https://arxiv.org/abs/2107.02075}{{\ttfamily 2107.02075}}.

\bibitem{Pasterski:2021rjz}
S.~Pasterski, \emph{{Lectures on celestial amplitudes}},
  \href{https://doi.org/10.1140/epjc/s10052-021-09846-7}{\emph{Eur. Phys. J. C}
  {\bfseries 81} (2021) 1062}
  [\href{https://arxiv.org/abs/2108.04801}{{\ttfamily 2108.04801}}].

\bibitem{Pate:2019lpp}
M.~Pate, A.-M. Raclariu, A.~Strominger and E.~Y. Yuan, \emph{{Celestial
  operator products of gluons and gravitons}},
  \href{https://doi.org/10.1142/S0129055X21400031}{\emph{Rev. Math. Phys.}
  {\bfseries 33} (2021) 2140003}
  [\href{https://arxiv.org/abs/1910.07424}{{\ttfamily 1910.07424}}].

\bibitem{Ball:2021tmb}
A.~Ball, S.~A. Narayanan, J.~Salzer and A.~Strominger, \emph{{Perturbatively
  exact w$_{1+\infty}$ asymptotic symmetry of quantum self-dual gravity}},
  \href{https://doi.org/10.1007/JHEP01(2022)114}{\emph{JHEP} {\bfseries 01}
  (2022) 114} [\href{https://arxiv.org/abs/2111.10392}{{\ttfamily
  2111.10392}}].

\bibitem{Melton:2022fsf}
W.~Melton, S.~A. Narayanan and A.~Strominger, \emph{{Deforming soft algebras
  for gauge theory}},
  \href{https://doi.org/10.1007/JHEP03(2023)233}{\emph{JHEP} {\bfseries 03}
  (2023) 233} [\href{https://arxiv.org/abs/2212.08643}{{\ttfamily
  2212.08643}}].

\bibitem{Melton:2023lnz}
W.~Melton and S.~A. Narayanan, \emph{{Celestial gluon amplitudes from the
  outside in}}, \href{https://doi.org/10.1007/JHEP05(2024)211}{\emph{JHEP}
  {\bfseries 05} (2024) 211}
  [\href{https://arxiv.org/abs/2312.12394}{{\ttfamily 2312.12394}}].

\bibitem{Kapec:2022axw}
D.~Kapec, Y.~T.~A. Law and S.~A. Narayanan, \emph{{Soft scalars and the
  geometry of the space of celestial conformal field theories}},
  \href{https://doi.org/10.1103/PhysRevD.107.046024}{\emph{Phys. Rev. D}
  {\bfseries 107} (2023) 046024}
  [\href{https://arxiv.org/abs/2205.10935}{{\ttfamily 2205.10935}}].

\bibitem{Kapec:2022hih}
D.~Kapec, \emph{{Soft particles and infinite-dimensional geometry}},
  \href{https://doi.org/10.1088/1361-6382/ad0514}{\emph{Class. Quant. Grav.}
  {\bfseries 41} (2024) 015001}
  [\href{https://arxiv.org/abs/2210.00606}{{\ttfamily 2210.00606}}].

\bibitem{Balthazar:2022hzb}
B.~Balthazar and C.~Cordova, \emph{{Geometry of conformal manifolds and the
  inversion formula}},
  \href{https://doi.org/10.1007/JHEP07(2023)205}{\emph{JHEP} {\bfseries 07}
  (2023) 205} [\href{https://arxiv.org/abs/2212.11186}{{\ttfamily
  2212.11186}}].

\bibitem{Cheung:2021yog}
C.~Cheung, A.~Helset and J.~Parra-Martinez, \emph{{Geometric soft theorems}},
  \href{https://doi.org/10.1007/JHEP04(2022)011}{\emph{JHEP} {\bfseries 04}
  (2022) 011} [\href{https://arxiv.org/abs/2111.03045}{{\ttfamily
  2111.03045}}].

\bibitem{Freidel:2022skz}
L.~Freidel, D.~Pranzetti and A.-M. Raclariu, \emph{{A discrete basis for
  celestial holography}},
  \href{https://doi.org/10.1007/JHEP02(2024)176}{\emph{JHEP} {\bfseries 02}
  (2024) 176} [\href{https://arxiv.org/abs/2212.12469}{{\ttfamily
  2212.12469}}].

\bibitem{Cotler:2023qwh}
J.~Cotler, N.~Miller and A.~Strominger, \emph{{An integer basis for celestial
  amplitudes}}, \href{https://doi.org/10.1007/JHEP08(2023)192}{\emph{JHEP}
  {\bfseries 08} (2023) 192}
  [\href{https://arxiv.org/abs/2302.04905}{{\ttfamily 2302.04905}}].

\bibitem{Guevara:2021abz}
A.~Guevara, E.~Himwich, M.~Pate and A.~Strominger, \emph{{Holographic symmetry
  algebras for gauge theory and gravity}},
  \href{https://doi.org/10.1007/JHEP11(2021)152}{\emph{JHEP} {\bfseries 11}
  (2021) 152} [\href{https://arxiv.org/abs/2103.03961}{{\ttfamily
  2103.03961}}].

\bibitem{Strominger:2021mtt}
A.~Strominger, \emph{{$w_{1+\infty}$ Algebra and the Celestial Sphere: Infinite
  Towers of Soft Graviton, Photon, and Gluon Symmetries}},
  \href{https://doi.org/10.1103/PhysRevLett.127.221601}{\emph{Phys. Rev. Lett.}
  {\bfseries 127} (2021) 221601}.

\bibitem{Himwich:2021dau}
E.~Himwich, M.~Pate and K.~Singh, \emph{{Celestial operator product expansions
  and w(1+infinity) symmetry for all spins}},
  \href{https://doi.org/10.1007/JHEP01(2022)080}{\emph{JHEP} {\bfseries 01}
  (2022) 080} [\href{https://arxiv.org/abs/2108.07763}{{\ttfamily
  2108.07763}}].

\bibitem{Atanasov:2021oyu}
A.~Atanasov, A.~Ball, W.~Melton, A.-M. Raclariu and A.~Strominger, \emph{{(2,
  2) Scattering and the celestial torus}},
  \href{https://doi.org/10.1007/JHEP07(2021)083}{\emph{JHEP} {\bfseries 07}
  (2021) 083} [\href{https://arxiv.org/abs/2101.09591}{{\ttfamily
  2101.09591}}].

\bibitem{Kutasov:1988xb}
D.~Kutasov, \emph{{Geometry on the Space of Conformal Field Theories and
  Contact Terms}},
  \href{https://doi.org/10.1016/0370-2693(89)90028-2}{\emph{Phys. Lett. B}
  {\bfseries 220} (1989) 153}.

\bibitem{Pasterski:2017kqt}
S.~Pasterski and S.-H. Shao, \emph{{Conformal basis for flat space
  amplitudes}}, \href{https://doi.org/10.1103/PhysRevD.96.065022}{\emph{Phys.
  Rev.} {\bfseries D96} (2017) 065022}
  [\href{https://arxiv.org/abs/1705.01027}{{\ttfamily 1705.01027}}].

\bibitem{Sharma:2021gcz}
A.~Sharma, \emph{{Ambidextrous light transforms for celestial amplitudes}},
  \href{https://doi.org/10.1007/JHEP01(2022)031}{\emph{JHEP} {\bfseries 01}
  (2022) 031} [\href{https://arxiv.org/abs/2107.06250}{{\ttfamily
  2107.06250}}].

\bibitem{De:2022gjn}
S.~De, Y.~Hu, A.~Yelleshpur~Srikant and A.~Volovich, \emph{{Correlators of four
  light-ray operators in CCFT}},
  \href{https://doi.org/10.1007/JHEP10(2022)170}{\emph{JHEP} {\bfseries 10}
  (2022) 170} [\href{https://arxiv.org/abs/2206.08875}{{\ttfamily
  2206.08875}}].

\bibitem{Hu:2022syq}
Y.~Hu, L.~Lippstreu, M.~Spradlin, A.~Y. Srikant and A.~Volovich,
  \emph{{Four-point correlators of light-ray operators in CCFT}},
  \href{https://doi.org/10.1007/JHEP07(2022)104}{\emph{JHEP} {\bfseries 07}
  (2022) 104} [\href{https://arxiv.org/abs/2203.04255}{{\ttfamily
  2203.04255}}].

\bibitem{Banerjee:2024hvb}
S.~Banerjee, R.~Basu and S.~Atul~Bhatkar, \emph{{Light transformation: A
  Celestial and Carrollian perspective}},
  \href{https://arxiv.org/abs/2407.08379}{{\ttfamily 2407.08379}}.

\bibitem{2018JHEP...11..102K}
P.~{Kravchuk} and D.~{Simmons-Duffin}, \emph{{Light-ray operators in conformal
  field theory}}, \href{https://doi.org/10.1007/JHEP11(2018)102}{\emph{Journal
  of High Energy Physics} {\bfseries 2018} (2018) 102}
  [\href{https://arxiv.org/abs/1805.00098}{{\ttfamily 1805.00098}}].

\bibitem{Donnay:2018neh}
L.~Donnay, A.~Puhm and A.~Strominger, \emph{{Conformally Soft Photons and
  Gravitons}}, \href{https://doi.org/10.1007/JHEP01(2019)184}{\emph{JHEP}
  {\bfseries 01} (2019) 184}
  [\href{https://arxiv.org/abs/1810.05219}{{\ttfamily 1810.05219}}].

\bibitem{Atanasov:2021cje}
A.~Atanasov, W.~Melton, A.-M. Raclariu and A.~Strominger, \emph{{Conformal
  block expansion in celestial CFT}},
  \href{https://doi.org/10.1103/PhysRevD.104.126033}{\emph{Phys. Rev. D}
  {\bfseries 104} (2021) 126033}
  [\href{https://arxiv.org/abs/2104.13432}{{\ttfamily 2104.13432}}].

\bibitem{Pasterski:2021fjn}
S.~Pasterski, A.~Puhm and E.~Trevisani, \emph{{Celestial diamonds: conformal
  multiplets in celestial CFT}},
  \href{https://doi.org/10.1007/JHEP11(2021)072}{\emph{JHEP} {\bfseries 11}
  (2021) 072} [\href{https://arxiv.org/abs/2105.03516}{{\ttfamily
  2105.03516}}].

\bibitem{Jorge-Diaz:2022dmy}
C.~Jorge-Diaz, S.~Pasterski and A.~Sharma, \emph{{Celestial amplitudes in an
  ambidextrous basis}},
  \href{https://doi.org/10.1007/JHEP02(2023)155}{\emph{JHEP} {\bfseries 02}
  (2023) 155} [\href{https://arxiv.org/abs/2212.00962}{{\ttfamily
  2212.00962}}].

\bibitem{Cheung:2022vnd}
C.~Cheung, A.~Helset and J.~Parra-Martinez, \emph{{Geometry-kinematics
  duality}}, \href{https://doi.org/10.1103/PhysRevD.106.045016}{\emph{Phys.
  Rev. D} {\bfseries 106} (2022) 045016}
  [\href{https://arxiv.org/abs/2202.06972}{{\ttfamily 2202.06972}}].

\bibitem{Dolan:2011dv}
F.~A. Dolan and H.~Osborn, \emph{{Conformal Partial Waves: Further Mathematical
  Results}},  \href{https://arxiv.org/abs/1108.6194}{{\ttfamily 1108.6194}}.

\end{thebibliography}\endgroup

\end{document}